\documentclass{aastex63}

\usepackage{graphics}
\usepackage{graphicx}

\received{}
\revised{}
\accepted{}

\begin{document}

\title{On the Alignment of Galaxies in Clusters}

\correspondingauthor{Hrant~M.~Tovmassian}
\email{htovmas@gmail.com}

\author{Hrant~M.~Tovmassian}
\affiliation{377, W. California 30\\
Glendale, CA, USA}

\author{J.P. Torres-Papaqui}
\affiliation{Departamento de Astronom\'ia, Universidad de Guanajuato\\
Apartado Postal 144, 36000, Guanajuato, Mexico}

\nocollaboration{2}

\begin{abstract}
We explore the distribution of position angles (PA) of galaxies in clusters. We selected for study the isolated 
clusters, since the distribution of the galaxy orientation in clusters with close neighbors could be altered by 
gravitational influence of the latter. We assume that galaxies are aligned, if their number at one $90^o$ 
position angle interval is more than twice higher than at the other $90^o$ interval. We study the galaxy PA 
distribution at the outer regions of clusters with smaller space density, where the probability of the PA variation 
in the result of interactions between galaxies is smaller than at the dense central regions. We found that the 
alignment of galaxies is more often observed in poor clusters and concluded that originally galaxies were 
aligned, but in the result of accretion in time of field galaxies with arbitrary orientations and also due to the 
mutual interactions the relative number of aligned galaxies decreases. 
\end{abstract}


\section{Introduction}

According to pancake scenario [1-3] galaxies form in the result of the gas-dust cloud collapse. At such case 
the angles (PA) of galaxies will naturally be aligned independent on the cluster mass. According to Miller \& 
Smith [4], Salvador-Sole \& Solanes [5] Usami \& Fujimoto [6], the galaxies could be aligned also at the 
hierarchical scenario of the cluster origin due to the cluster tidal field. At the latter scenario the galaxies will be 
aligned predominantly in rich clusters. Thus the alignment of galaxies in clusters is a clue for explanation of 
their origin. Therefore a lot of efforts have been undertaken in the past for study the distribution of the position 
angles (PA) of galaxies in clusters.
Some evidence on the alignment of galaxies with the parent cluster were reported by Sastry [7] Adams, Strom 
\& Strom [8], Carter \& Metcalfe [9], Binggeli [10], Struble \& Peebles [11], Rhee \& Katgert [12], Lambas et al. 
[13], Flin \& Olowin [14], Fong, Stevenson \& Shanks [15]. More certainly the alignment was found between 
orientations of the cluster and of the BCG (cD) [7-9, 16-22]. Plionis et al.[23] and Rong,  Zhang \& Liao [24] 
found an evidence that significant galaxy alignment is present in dynamically young clusters. Meanwhile, 
Dekel [25], van Kampen \& Rhee [26], Trevese, Cirimele \& Flin 27], Djorgovski [28] and Cabanela 
\& Aldering [29] found no galaxy alignment, except the alignment of the BCG with its parent cluster. Chen et 
al. [30] found a statistically significant galaxy-filament correlation, but not on the galaxy-cluster alignment.
Thus, the results on the study of the galaxy alignment in clusters were contradictory. 

In this paper we undertook new search for alignment of galaxies and showed that galaxies are aligned mostly 
in poor clusters. We suggest that the primordial orientations of member galaxies were ordered at the cluster 
origin, but later on the assembly of field galaxies by the cluster and interactions between galaxies within the 
cluster introduce disorder in the galaxy orientations. We showed also that clusters in which the primordial 
alignment of galaxies preserved, do rotate.

\section{The data}

We study the possible alignment of galaxies in ACO [31] clusters. Many ACO clusters are themselves 
clustered [32-34]. The gravitational influence of the nearby cluster may affect on the orientation of galaxies in 
the studied cluster. In order to avoid this effect we studied isolated clusters. We compiled a list of 73 strongly 
isolated clusters with nearest neighbor located on sky at the projected distance $>10$~Mpc (Table 1). 
For comparison we studied also the clusters with smaller degree of isolation, with 5 to 9 Mpc 
projected distance to the nearest neighbor on sky. The list of the mild isolated 25 clusters is presented in 
Table \ref{Table02}. Redshifts of the selected clusters are $z<0.1$ and they contain more than 20 galaxies 
within area with 2 Mpc radius. It is assumed that member galaxies of the most ACO clusters are located within 
2 Mpc of the Abell radius [32], defined as $R_A =1\arcmin .7/z$ (H$_0$ = 72 $km\, s^{-1} Mpc^{-1}$) [35].
The member galaxies of clusters were retrieved from the SDSS-DR9 [36]. The galaxies with the primary mode 
(marked in the catalog  as 1) and good quality of observations (marked by 3) were retrieved. According to  
[37] we retrieved galaxies with velocities within $\pm1500$ km s$^{-1}$ of the cluster velocity. PAs are those 
at $r$ band. 

At the corresponding columns of Tables 1 and 2 the following information is given: the cluster designation, the 
redshift of the cluster and the number of galaxies in the cluster within 2 Mpc radius. The parameters of clusters 
are from NED.\footnote {The NASA/IPAC 
Extragalactic Database (NED) is operated by the Jet Propulsion Laboratory, California Institute of Technology, 
under contract with the National Aeronautics and Space Administration.}

\begin{table*}
\caption{The list of isolated clusters with nearby neighbor at projected distance $>10$ Mpc.}
\label{Table 1}
\begin{tabular}{llllll}
\hline\hline
Cluster &   $z$  & $N_2$ & Cluster &   $z$  & $N_2$      \\
\hline
A595	&	0.0666	&	48	&  A1552	&	0.0858	&	74	\\
A602	&	0.0619	&	62	&  A1564	&	0.0792	&	52	\\
A634	&	0.0265	&	102	&  A1599	&	0.0855	&	25	\\
A635	&	0.0925	&	34	&  A1609	&	0.0891	&	27	\\
A660	&	0.0642	&	25	&  A1616	&	0.0833	&	48	\\
A671	&	0.0502	&	98	&  A1630	&	0.0648	&	36	\\
A690	&	0.0788	&	49	&  A1684	&	0.0862	&	27	\\
A692	&	0.0894	&	50	&  A1692	&	0.0842	&	49	\\
A695	&	0.0687	&	27	&  A1750	&	0.0852	&	91	\\
A699	&	0.0851	&	31	&  A1781	&	0.0618	&	45	\\
A724	&	0.0933	&	46	&  A1783	&	0.0690	&	50	\\
A727	&	0.0951	&	58	&  A1809	&	0.0791	&	94	\\
A744	&	0.0729	&	32	&  A1812	&	0.0630	&	28	\\
A757	&	0.0517	&	49	&  A1825	&	0.0595	&	30	\\
A779	&	0.0225	&	115	&  A1827	&	0.0654	&	41	\\
A819	&	0.0759	&	20	&  A1849	&	0.0963	&	27	\\
A834	&	0.0709	&	35	&  A1864	&	0.0870	&	51	\\
A858	&	0.0863	&	26	&  A1890	&	0.0574	&	83	\\
A1024	&	0.0734	&	49	&  A2018	&	0.0878	&	50	\\
A1028	&	0.0908	&	26	&  A2019	&	0.0807	&	24	\\
A1035	&	0.0684	&	59	&  A2022	&	0.0578	&	78	\\
A1066	&	0.0690	&	83	&  A2048	&	0.0972	&	61	\\
A1126	&	0.0646	&	33	&  A2082	&	0.0862	&	24	\\
A1139	&	0.0398	&	50	&  A2107	&	0.0411	&	130	\\
A1142	&	0.0349	&	64	&  A2108	&	0.0919	&	48	\\
A1168	&	0.0906	&	41	&  A2110	&	0.0980	&	27	\\
A1169	&	0.0586	&	79	&  A2122	&	0.0661	&	72	\\
A1238	&	0.0733	&	68	&  A2142	&	0.0909	&	123	\\
A1270	&	0.0692	&	63	&  A2148	&	0.0877	&	30	\\
A1307	&	0.0817	&	67	&  A2162	&	0.0322	&	47	\\
A1314	&	0.0335	&	119	&  A2178	&	0.0928	&	24	\\
A1371	&	0.0398	&	61	&  A2205	&	0.0876	&	39	\\
A1424	&	0.0768	&	72	&  A2255	&	0.0806	&	122	\\
A1480	&	0.0734	&	31	&  A2366	&	0.0529	&	53	\\
A1507	&	0.0604	&	57	&  A2593	&	0.0413	&	138	\\
A1516	&	0.0769	&	60	&  A2630	&	0.0667	&	37	\\
A1541	&	0.0893	&	79	& 						\\
\hline\hline
\end{tabular}
\end{table*}

\begin{table*}
\caption{The list of mild isolated clusters with nearby neighbor at projected distance $5<d<9$~Mpc.}
\label{Table 2}
\begin{tabular}{llllll}
\hline\hline
Cluster &  $z$  & $N_2$ & Cluster &  $z$  & $N_2$  \\
\hline
\hline
A912  &  0.0446  & 21 	&  A1691 &  0.0721  & 75  \\
A933  &  0.0956  & 47 	&  A1775 &  0.0717  & 60  \\
A1100 &  0.0463  & 53 &  A1795 &  0.0625  & 103 \\
A1149 &  0.0710  & 37 &  A1831 &  0.0615  & 37  \\
A1185 &  0.0325  & 182	&  A1927 &  0.0945  & 35  \\
A1205 &  0.0754  & 46 	&  A1983 &  0.0436  & 149 \\
A1267 &  0.0329  & 28 	&  A1991 &  0.0587  & 99  \\
A1291 &  0.0527  & 84 	&  A2029 &  0.0773  & 77  \\
A1468 &  0.0844  & 31 	&  A2065 &  0.0726  & 115 \\
A1589 &  0.0725  & 100	&  A2089 &  0.0711  & 60  \\
A1638 &  0.0620  & 33 	&  A2092 &  0.0669  & 50  \\
A1650 &  0.0838  & 59 	&  A2149 &  0.0679  & 40  \\
A1663 &  0.0843  & 60 	&				\\
\hline\hline
\end{tabular}
\end{table*} 

\section{Analysis}

We used a simple method for search of the alignment of the orientation of galaxies in clusters. We divided the 
range of PAs of galaxies in each cluster into two $90^o$ sections so that to have high number ($N_h$) of 
galaxies at one section and small number ($N_s$) of galaxies at the other section. We assume that there is 
an alignment signal, if the number of galaxies at one $90^o$ section is by at least 2 times higher than at the 
other $90^o$ section. 

A primordial galaxy alignments in clusters could be severely damped by the violent relaxation, by the 
exchange of angular momentum in galaxy interactions over a Hubble time [38] that mostly occur in the dense 
cluster environment. Therefore, we first searched the orientation of galaxies in the outer area of clusters at 
the ring with cluster-centric radii $1\div2$ Mpc.

The results of counts in the outer ring of clusters with smaller degree of isolation is presented in Table 4. In the 
corresponding columns of Table 3 the following information is given: 1st - the 
cluster designation; 2d - the interval of PAs at which the high number of galaxies are distributed; 3d - the 
number $N_h$ of galaxies at this section; 4th - the number of galaxies at the opposite section; 5th - the ratio 
$N_h/N_s$ at the searched region. In these clusters the alignment signal was found only for 8 out of 26 
clusters, 32\%.

At the ring with cluster-centric radii $1\div$ 2 Mpc 43 clusters out of 73, i.e 59\% have alignment signal 
(Table 3). Note that in the case of a random distribution of PAs the numbers of galaxies in two $90^o$ 
intervals could occasionally differ from each other by more than 2 times. In order to verify whether the the 
found number of clusters with alignment signal are real or are a result of random distribution of the galaxy PAs 
we applied non-parametric bootstrapping statistical test making 1000 simulations. The same statistical test 
was applies below for checking the reality of the found alignments in other cluster samples.

The probability that in 43 clusters out of 73 the ratio $N_h/N_s$ exceeds 2 is real and is not a result of a 
random distribution is 58.90\% of success and a 95 percent confidence interval from 46.76\% to 70.29\% 
with a p-value = 0.01597. The p-value or probability value (>0.05) is the probability of obtaining test results 
at least as extreme, as the results actually observed during the test, assuming that the null hypothesis is 
correct. Hence, the probability that the found galaxy alignments at rings are real, is sufficiently 
high.

\begin{table*}
\caption{The list of strongly isolated clusters with alignment signal for galaxies at the ring with radii 
$1\div2$ Mpc.}
\label{Table 3}
\begin{tabular}{lllll}
\hline\hline
Cluster &   $I_{PA}$ &  $N_h$ & $N_s$ & $\left(\frac{N_h}{N_s}\right)_{1\div 2}$  \\
\hline
	&       degree    			  \\
\hline
\hline
A595	&	37-122	&	12	&	3	&	4.0	  \\
A635A	&	78-165	&	8	&	2	&	4.0	  \\
A660A	&	160-69	&	12	&	4	&	2.2	  \\
A695A	&	26-107	&	9	&	3	&	3.0	  \\
A699A	&	82-158	&	12	&	6	&	2.0	  \\
A724	&	171-81	&	14	&	3	&	4.7	  \\
A744A	&	2-86	&	7	&	3	&	2.3	  \\
A819A	&	23-113	&	7	&	3	&	2.3	  \\
A834	&	18-106	&	15	&	4	&	3.75	  \\
A858	&	56-136	&	8	&	2	&	4.0	  \\
A1028A	&	  1-89	&	9	&	4	&	2.25	  \\
A1035A	&	69-158	&	8	&	4	&	2.0	  \\
A1126A	&	18-108	&	11	&	5	&	2.2	  \\
A1139	&	  6-96	&	12	&	6	&	2.0	  \\
A1168	&	  3-84	&	12	&	5	&	2.4	  \\
A1169	&	20-108	&	23	&	8	&	2.9	  \\
A1238A	&	102-10	&	23	&	7	&	3.3	  \\
A1270	&	69-159	&	17	&	7	&	2.4	  \\
A1307	&	164-74	&	24	&	11	&	2.2	  \\
A1371	&	39-121	&	18	&	7	&	2.6	  \\
A1480A	&	161-70	&	11	&	5	&	2.2	  \\
A1516A	&	76-160	&	18	&	8	&	2.25	  \\
A1541	&	64-144	&	25	&	7	&	3.6	  \\
A1552	&	131-41	&	24	&	12	&	2.0	  \\
A1599A	&	100-3	&	7	&	3	&	2.3	  \\
A1616A	&	2-89	&	16	&	6	&	2.7	  \\
A1684A	&	68-156	&	8	&	4	&	2.0	  \\
A1781A	&	68-156	&	19	&	7	&	2.7	  \\
A1783A	&	88-167	&	12	&	3	&	4.0	  \\
A1809A	&	15-103	&	24	&	12	&	2.0	  \\
A1812A	&	32-122	&	12	&	4	&	3.0	  \\
A1825A	&	115-20	&	11	&	5	&	2.2	  \\
A1849A	&	73-150	&	9	&	2	&	4.5	  \\
A2018A	&	44-131	&	21	&	7	&	2.6	  \\
A2019A	&	75-156	&	9	&	4	&	2.25	  \\
A2082A	&	144-40	&	11	&	1	&	11.0	  \\
A2108	&	32-108	&	12	&	3	&	4	  \\
A2110	&	91-180	&	11	&	3	&	3.7	  \\
A2122A	&	51-141	&	22	&	11	&	2.0	  \\
A2148	&	108-18	&	8	&	3	&	2.3	  \\
A2178A	&	51-133	&	9	&	4	&	2.25	  \\
A2366A	&	85-170	&	9	&	4	&	2.25	  \\
A2630A	&	44-132	&	10	&	2	&	5	  \\
\hline\hline
\end{tabular}
\end{table*}

The orientations of 19 clusters with alignment signal studied in this paper were determined by Plionis [19]. In 
Figure 1 the distribution of PAs of galaxies in these clusters and the PAs of the cluster large axis 
determined by Plionis [19] are shown. The PAs of the large axes of 14 clusters are within interval of PAs of 
aligned galaxies. For 2 clusters, A1783 and A1812, the PA of their large axes fall into the interval of PAs of 
aligned galaxies, if to take into account the errors of the PA measurements [19] about $30^o$. The PAs of 
large axes of only 3 clusters, A1126, A1139, and A1812 are out of the $90^o$ interval of the PAs of galaxies 
with alignment signal. The probability of 16 chance coincidences out of 19 is sufficiently small, 0.01. Even with 
exclusion of A1783 and A1812, the probability of the chance coincidences of 14 out of 19 is still small, 0.02. 
The coincidence of the cluster large axes orientation with the interval of PAs of the majority of the cluster 
galaxies shows that the applied simple method for searching the alignment of galaxies in clusters is reliable.

\begin{figure} 
\centering
\includegraphics[width=12cm]{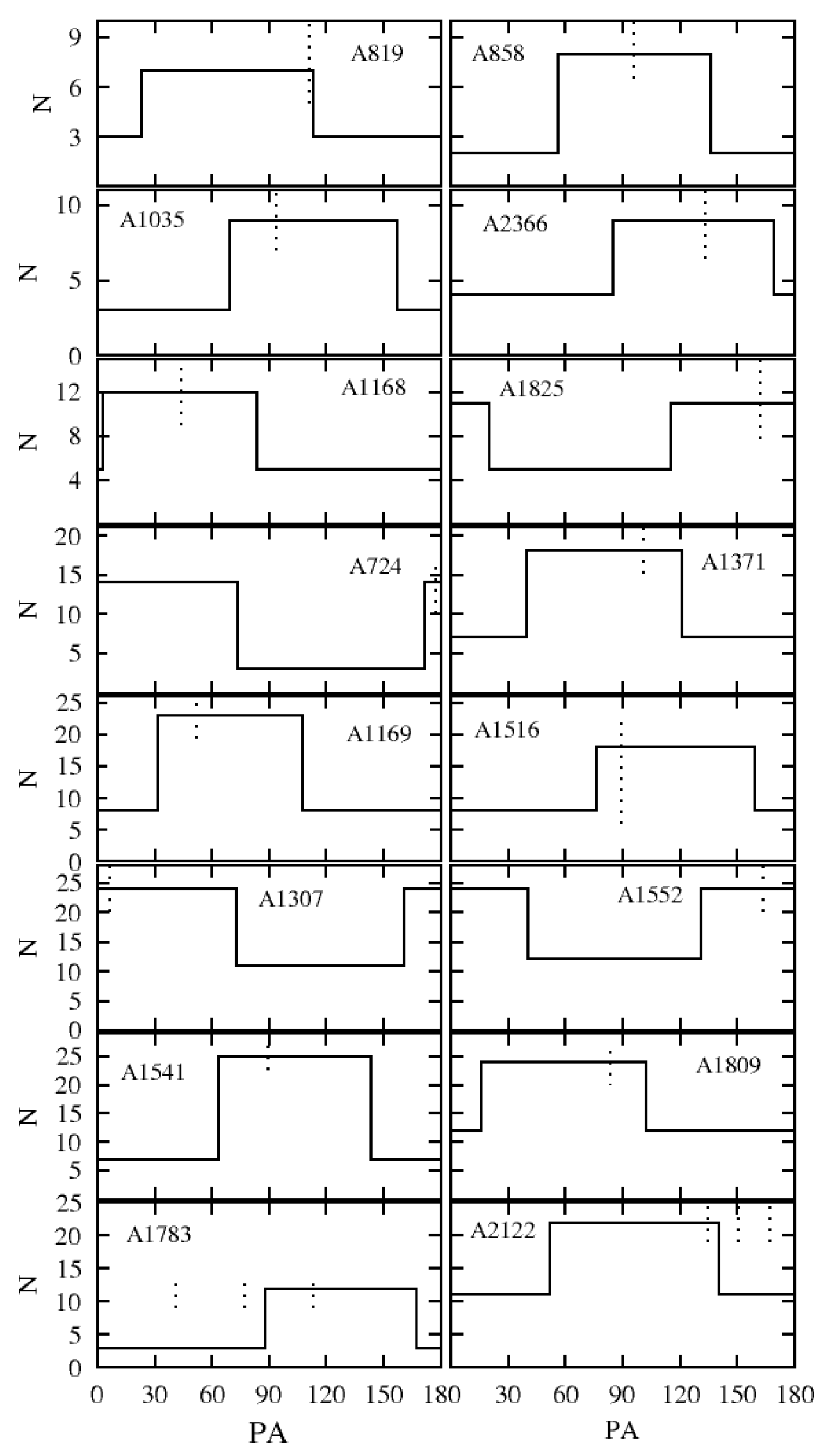}
\caption{The comparison of the distribution of PAs of galaxies in clusters with alignment signal with the cluster 
PA from Plionis [19] shown by dotted line. In cases of A1783 and A2122 the interval of the errors of the PA of 
the large axes of the cluster are also shown.}
\label{Figure 1}
\end{figure}

The results of counts in the outer ring of clusters with smaller degree of isolation and at the central area of 
strongly isolated clusters are presented respectively in Table 4 and Table 5 identical to 
Table 1.

\begin{table*}
\caption{The list of the mild isolated clusters with alignment signal for galaxies at the ring with radii $1\div2$ 
Mpc.}
\label{Table 4}
\begin{tabular}{lllll}
\hline\hline
Cluster &   $I_{PA}$ &   $N_h$ & $N_s$ & $\left(\frac{N_h}{N_s}\right)_{1\div 2}$  \\
\hline
       &    degree     &    			  \\
\hline
\hline
A933  &  24-111  & 18 &	8	&	2.25  \\
A1100 &  83-180  & 15 &	7	&	2.1   \\
A1119 &  63-157  & 8  &	4	&	2.0   \\
A1205 &  96-171  & 8  &	4	&	2.0   \\
A1267 &  43-140  & 13 &	4	&	3.25  \\
A1468 & 124-6    & 9  &	4	&	2.25  \\
A1775 &  45-132  & 37 &	17	&	2.2   \\
A1927 &  57-141  & 17 &	4	& 	4.25  \\
A2149 &  16-104  & 12 &	4	&	3.0   \\
\hline\hline
\end{tabular}
\end{table*}

\begin{table*}
\caption{ The list of clusters with aligned galaxies at the cluster central region.}
\label{Table 5}
\begin{tabular}{lllll}
\hline\hline
Cluster &   $I_{PA}$ &   $N_h$ & $N_s$ & $\left(\frac{N_h}{N_s}\right)_1$   \\
	& degree	&		&		&	     \\
\hline
\hline
A595	&	37-122	&	12	&	3	&	2.1  \\
A660A	&	160-69	&	12	&	4	&	2.0  \\
A695A	&	26-107	&	9	&	3	&	2.7  \\
A727	&	16-106	&	22	&	17	&	2.8  \\
A744A	&	 2-86	&	7	&	3	&	4.5  \\
A834	&	18-106	&	15	&	4	&	3.0  \\
A1028A	&	  1-89	&	9	&	4	&	2.25 \\
A1126A	&	18-108	&	11	&	5	&	3.0  \\
A1168	&	  3-84	&	12	&	5	&	2.7  \\
A1238A	&	102-10	&	23	&	7	&	2.75 \\
A1371	&	39-121	&	18	&	7	&	2.0  \\
A1541	&	64-144	&	25	&	7	&	2.5  \\
A1552	&	131-41	&	24	&	12	&	2.2  \\
A1564A	&	32-122	&	13	&	12	&	2.25 \\
A1599A	&	100-3	&	7	&	3	&	2.0  \\
A1616A	&	  2-89	&	16	&	6	&	2.7  \\
A1630A	&	7-96	&	7	&	4	&	2.6  \\
A1692	&	9-96  	&	17	&	9	&	3.0  \\
A1750A	&	13-103	&	36	&	23	&	2.7  \\
A1781A	&	68-156	&	19	&	7	&	2.1  \\
A1812A	&	32-122	&	12	&	4	&	3.0  \\
A2019A	&	75-156	&	9	&	4	&	2.7  \\
A2048	&	  0-90	&	19	&	14	&	2.1  \\
A2082A	&	144-40	&	11	&	1	&	3.0  \\
A2108	&	32-108	&	12	&	3	&	3.1  \\
A2122A	&	51-141	&	22	&	11	&	2.2  \\
A2205A	&	46-136	&	14	&	10	&	2.0  \\
\hline\hline
\end{tabular}
\end{table*}

In the outer ring of clusters with smaller degree of isolation the alignment signal is found only for 8 out of 25 
clusters, 32\%. The probability that the found alignment signal in the ring of 8 out of 25 clusters is real, is 
32.00\% of success with a 95 percent confidence interval from 14.94\% to 53.50\% with a p-value = 0.01078. 

At the central region of strongly isolated clusters the alignment signal is found for 27 
clusters, 36\%, out of the studied 73. The probability that the alignment at the central area of 
clusters does not occur by chance and is real is 36.98\% of success and a 95 percent confidence interval from 
25.97\% to 49.08\% with a p-value = 0.01442. Hence, the found alignment of galaxies in both cluster samples 
are caused rather by random distribution of the galaxy PAs and are not real.

\subsection{The dependence of the alignment on the cluster richness}

The alignment of galaxies could depend on the richness of clusters and/or on the absolute magnitude of the 
observed galaxies, i.e. on the cluster redshift. In order to find out whether the alignment signal depends on the 
cluster richness or distance we split the list of 73 strongly isolated clusters into two parts: with low and high 
redshift clusters, and also poor and rich ones. In the consecutive lanes of Table 6 the average 
redshift $z$, the average number $N_2$ of galaxies, the minimal absolute magnitude $<M_r>$ in $r$-band 
for the average redshift, the number $N_{as}$ of galaxies with alignment signal and the ratio $N_{as}/N_t$ of 
the number of clusters with alignment signal to the total number of clusters are presented for nearby (column 
2) and distant (column 3) clusters. In Table 7 the same data are presented for samples of rich and 
poor clusters.

\begin{table}
\caption{The parameters of the nearby and distant clusters.}
\label{Table 6}
\begin{tabular}{rrr}
\hline\hline
	& Nearby clusters & Distant clusters  \\
\hline
$<z>$		& 0.0565$\pm$0.0143	& 0.0860$\pm$0.0066   \\
$<N_2>$		&	62$\pm$30	&	50$\pm$26	\\
$<M_r>$		&	-19.09		&	-20.02		\\
$N_{as}$	&	19		&	 24		\\
$N_{as}/N_t$	& 0.51$\pm$0.12	& 0.65$\pm$0.13	\\
\hline\hline
\end{tabular}
\end{table}

\begin{table}
\caption{The parameters of poor and rich clusters.}
\label{Table 7}
\begin{tabular}{rrr}
\hline\hline
	& Rich clusters  &   Poor clusters  \\
\hline
$z$		& 0.0655$\pm$0.0207	& 0.0774$\pm$0.0143   \\
$<N_2>$		&	77$\pm$25	&	35$\pm$9	\\
$<M_r>$		&	-19.47		&	-19.78		\\
$N_{as}$	&	15		&	28		\\
$N_{as}/N_t$	& 0.40$\pm$0.10	& 0.76$\pm$0.14	\\
\hline\hline
\end{tabular}
\end{table}

Table 6 shows that difference between relative numbers of galaxies with alignment signal in 
nearby and distant clusters is not high, although the distant clusters are on average by 1.5 times farther and 
the limiting absolute $M_r$ magnitudes of galaxies in this clusters differ by about $1^m$. The average total 
numbers of galaxies in clusters of both samples also does not differ from each other significantly. The relative 
number of nearby clusters with alignment signal is 0.51 with 50.06\% of success and 95 percent confidence 
interval from 38.71\% to 62.59\% with a p-value = 3.182e-03. The relative number of distant clusters with 
alignment signal is 0.65, with 64.38\% of success and a 95 percent confidence interval from 52.30\% to 
75.25\% with a p-value = 1.818e-04.

The situation is different when we compare rich and poor clusters. Table 7 shows that the 
differences between the average redshifts and the limiting absolute magnitudes $M_r$ of these two samples 
are smaller in comparison to those in nearby and distant clusters. However, the relative number of poor 
clusters with alignment signal is by about 2.6 times higher in comparison to rich ones. The relative number of 
rich clusters with alignment signal is 0.40 with 39.72\% of success and a 95 percent confidence interval from 
28.45\% to 51.85\% with a p-value = 1.173e-02. The relative number of poor clusters with alignment signal 
is 0.76 with 75.34\% of success and a 95 percent confidence interval from 63.85\% to 84.68\% with a 
p-value = 1.514e-085. Thus, in poor clusters the probability of the reality of the found alignment is 
sufficiently high, about 80\%.

\section{Discussion and Conclusions}

By study of the distribution of PAs of galaxies in the ring with radii $1\div2$ Mps of 73 strongly isolated 
clusters the alignment signal is found in 43 clusters, i.e. in about 60\%. Such high number of clusters with 
aligned galaxies may not be caused by a chance distribution of the galaxy PAs. Among the less isolated 
clusters and in the central dense area of clusters with 1 Mpc radius the alignment signal is found respectively 
in about 37\% and 29\% of clusters, that is close to the expected number of a chance occurrence of the 
alignment signal, the ratio $N_h/N_s>2$. The separate analysis of clusters of different richnesses and 
distances showed that the alignment depends on the cluster richness. Alignment is found in about 75\% of 
poor clusters with on average 35 galaxies within 2~Mpc. The probability that this is not due by random 
distribution of the galaxy PAs is suficiently high. This evidences in favor of the pancake scenario [1,2] of the 
cluster formation. If so, clusters could preserve the angular momentum of the primordial gas cloud. 

According to Miller \& Smith [4], Salvador-Sole \& Solanes [5] Usami \& Fujimoto [6], the galaxies 
could as well be aligned in the hierarchical scenario due to the tidal field of the cluster. However, the tidal field 
of the cluster would apparently be more effective in rich clusters with higher mass and the alignment would be 
observed in rich clusters. Whereas, we found the opposite.

During the cluster evolution the primordial alignment of galaxies could be altered. The alignment rate will 
decrease in the result of gravitational influence of nearby clusters and mutual interactions between galaxies. 
Apparently the rate of interactions is higher in rich clusters and especially at the cluster dense central regions. 
The gravitational influence would apparently have smaller effect on the orientation of massive galaxies. 
Therefore, the alignment of only very massive BCGs (cDs) has been found with the cluster orientation
[7-10, 17-22]. The inclusion to the cluster content of the faint field galaxies by the hierarchical assembly 
[39 and references therein] with arbitrary orientations will certainly decrease the 
relative number of aligned galaxies. The poorer is the cluster, i.e. the less massive it is, the smaller amount of 
field galaxies would be assembled. Thus, the primordial alignment is better preserved in poor clusters, in 
which both reasons for altering it, interactions between galaxies and assembly of the field galaxies are less 
effective.

\acknowledgments
We are grateful to M. Plionis for presentation the list of clusters with their neighbors and to the anonymous 
referee for careful reading of the manuscript and valuable comments. T-P acknowledges for 
support through grant DAIP-UGto (0173/19). This research has made use of the NASA/IPAC Extragalactic 
Database (NED), which is operated by the Jet Propulsion Laboratory, California Institute of Technology, under 
contract with the National Aeronautics and Space Administration. Funding for SDSS-III has been provided by 
the Alfred P. Sloan Foundation, the Participating Institutions, the National Science Foundation, and the U.S. 
Department of Energy Office of Science. The SDSS-III web site is http://www.sdss3.org/.

SDSS is managed by the Astrophysical Research Consortium for the Participating Institutions of the SDSS-
III Collaboration including the University of Arizona, the Brazilian Participation Group, Brookhaven National 
Laboratory, Carnegie Mellon University, University of Florida, the French Participation Group, the German 
Participation Group, Harvard University, the Instituto de Astrofisica de Canarias, the Michigan State/Notre 
Dame/JINA Participation Group, JohnsHopkins Univirsity, Lawrence Berkeley National Laboratory, Max 
Planck Institute for Astrophysics, Max Planck Institute for Extraterrestrial Physics, New Mexico State 
University, New York University, Ohio State University, Pennsylvania State University, University of 
Portsmouth, Princeton University, the Spanish Participation Group, University of Tokyo, University of Utah, 
Vanderbilt University, University of Virginia, University of Washington, and Yale University.

\end{document}